\documentclass[conference]{IEEEtran}
\IEEEoverridecommandlockouts
\usepackage{cite}
\usepackage{amsmath,amssymb,amsfonts}
\usepackage{algorithmic}
\usepackage{graphicx}
\usepackage{textcomp}
\usepackage{xcolor}
\usepackage{orcidlink}
\usepackage{comment}

\usepackage{fancyhdr} 
\def\BibTeX{{\rm B\kern-.05em{\sc i\kern-.025em b}\kern-.08em
    T\kern-.1667em\lower.7ex\hbox{E}\kern-.125emX}}
\begin{document}

\title{HARP: A Large-Scale Higher-Order Ambisonic Room Impulse Response Dataset}


\author{\IEEEauthorblockN{Shivam Saini\orcidlink{0009-0002-3981-2260}}
\IEEEauthorblockA{\textit{Institut für Kommunikationstechnik} \\
\textit{Leibniz University Hannover}\\
Hannover, Germany \\}
\and
\IEEEauthorblockN{Jürgen Peissig}
\IEEEauthorblockA{\textit{Institut für Kommunikationstechnik} \\
\textit{Leibniz University Hannover}\\
Hannover, Germany \\}
}

\maketitle

\begin{abstract}
This contribution introduces a dataset of 7th-order Ambisonic Room Impulse Responses (HOA-RIRs), created using the Image Source Method. By employing higher-order Ambisonics, our dataset enables precise spatial audio reproduction, a critical requirement for realistic immersive audio applications. Leveraging the virtual simulation, we present a unique microphone configuration, based on the superposition principle, designed to optimize sound field coverage while addressing the limitations of traditional microphone arrays. The presented 64-microphone configuration allows us to capture RIRs directly in the Spherical Harmonics domain. The dataset features a wide range of room configurations, encompassing variations in room geometry, acoustic absorption materials, and source-receiver distances. A detailed description of the simulation setup is provided alongside for an accurate reproduction. The dataset serves as a vital resource for researchers working on spatial audio, particularly in applications involving machine learning to improve room acoustics modeling and sound field synthesis. It further provides a very high level of spatial resolution and realism crucial for tasks such as source localization, reverberation prediction, and immersive sound reproduction.
\end{abstract}


\section{Introduction}

Spatial audio reproduction plays a vital role in enhancing user immersion in modern applications such as virtual reality (VR), augmented reality (AR), and interactive environments. The development of precise methods for capturing and reproducing sound fields is essential for accurately conveying the spatial characteristics of acoustic environments. Higher-order ambisonics (HOA) has emerged as a leading technique for this purpose, allowing for increased spatial resolution and accurate 3D audio rendering through spherical harmonic encoding \cite{moreau20063d}.

However, the availability of high-quality, large-scale datasets specifically designed for HOA-based room impulse responses (HOA-RIRs) is limited. Most current datasets either focus on first-order Ambisonics failing to capture the complex sound interactions in real-world environments or does not contain enough variation of rooms and configurations for statistical analysis and machine learning based tasks. To address these limitations, we present a dataset of 7th-order HOA RIRs, generated using the Image Source Method (ISM), which simulates relatively more realistic room acoustics for a variety of room configurations.

This paper describes the methodology for generating the proposed dataset HARP, including the microphone arrangement, simulation parameters, and spherical harmonic decomposition techniques used. The dataset provides an essential resource for researchers working in spatial audio, room acoustics, and machine learning, enabling the development of new algorithms for sound field analysis and immersive audio reproduction. We provide a rich, diverse resource for training and evaluating machine learning models that look over real-world reverberation applications such as Room parameter estimation \cite{AudMobNet, bryanCNN, CRNN, gotzjournal}, Dereverberation \cite{habets2007single},  Immersive Music Generation \cite{jin2022metamgc}, Spatial Audio Compression \cite{hold2024}, Reverberation Synthesis \cite{nilsI3Dparameteric}, Spatial Upsampling \cite{xia2023upmix}, Sound Event Localization and Detection Tasks \cite{politis2020overview} etc. One major contribution of this research is the implementation of a $7^{th}$-Order Ambisonics microphone configuration in \texttt{Pyroomacoustics} library\cite{pyroomacoustics}.

\section{Related Work}

Room Impulse Responses (RIRs) are fundamental to room acoustics, providing a detailed characterization of how sound propagates within an enclosed space. Traditional methods of RIR generation often rely on either measured responses in real rooms or computational methods such as the image-source method or ray tracing. These approaches have been used extensively in both acoustic modelling and spatial audio applications. 

On the other hand, Higher-order Ambisonics (HOA) has gained significant traction in recent years as a means of capturing and reproducing 3D audio scenes \cite{zotter2019ambisonics}. HOA with the use of Spherical Harmonics (SH) provides a more detailed description of the sound field, making it more useful when rendering for immersive applications \cite{frank2014}. However, there's a shortage of existing HOA Impulse Responses datasets due to the computational cost and efforts involved. 

Most existing multi-channel datasets, however, are limited in scale (few room configurations) and often employ microphone arrays that do not provide optimal spatial coverage. FOr instance, traditional ambisonics datasets are capturedd using first-order microphones, which can lead to inaccuracies. This limitation arises from the finite number of transducers used to reproduce a physical sound field, resulting in issues such as localization blur \cite{bertet2013}, reduced lateralization \cite{daniel1998ambisonics}, and comb-filtering spectral artefacts \cite{jot1999comparative}. Our approach addresses this gap by generating a dataset of synthetic 7th-order Ambisonic RIRs using superpositioned microphone placement which ensures uniform spatial sampling and enhances the accuracy of spherical harmonic-based sound field encoding.

To our knowledge, this is the first large-scale HOA-RIR dataset, contributing both a novel dataset and a new method for spatial sampling in the context of room acoustics. Our dataset aims to support a variety of signal processing applications, including spatial audio rendering, room acoustics modelling, and machine learning-based sound field reconstruction. A comparison to the existing RIR dataset is given later in Discussion and Table \ref{tab:dataset_comparison}.

\section{Methodology}

\begin{figure}[!h]
    \centering
    \includegraphics[width=0.95\columnwidth]{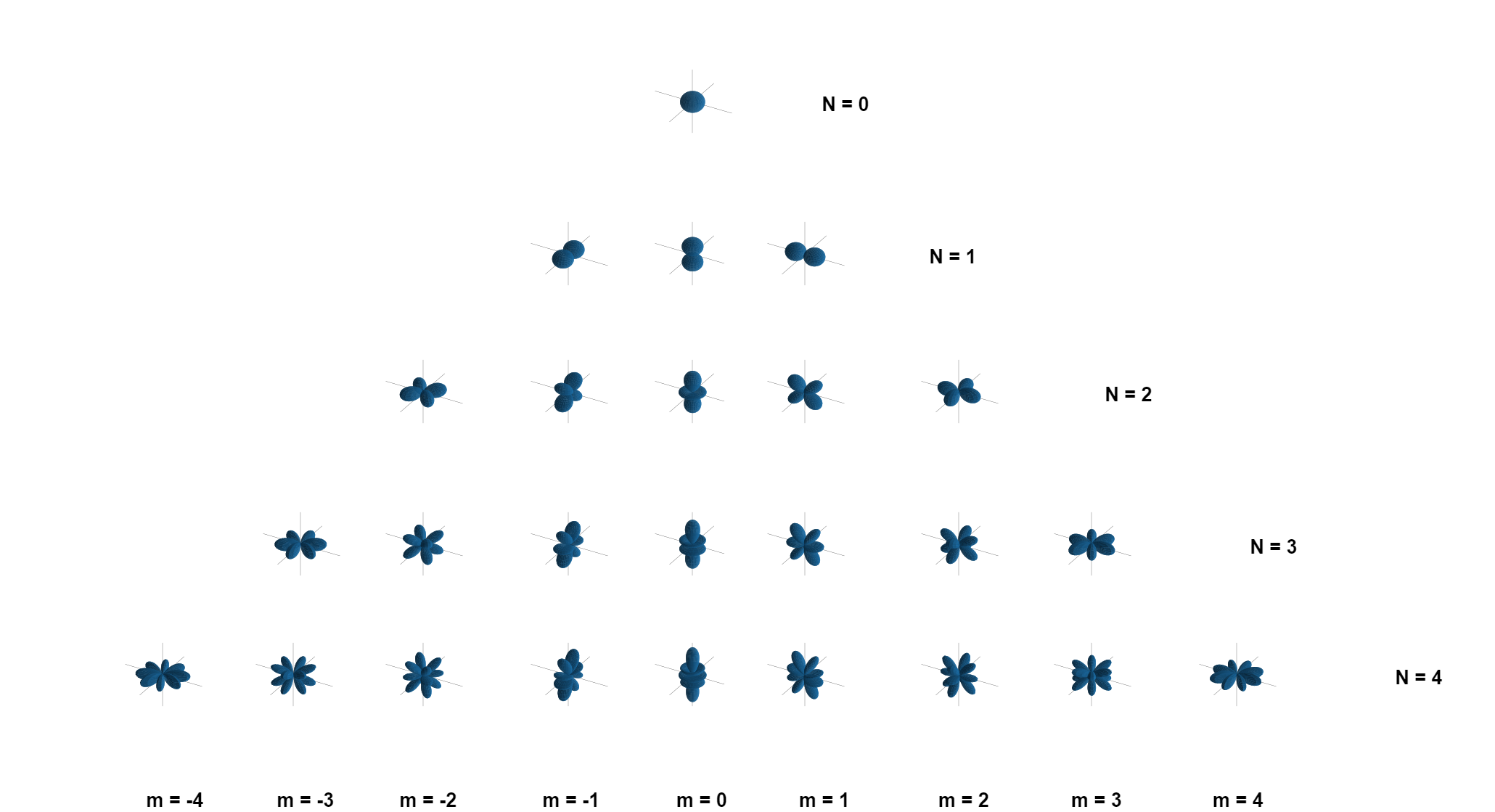}
    \caption{Spherical harmonics up to 4th order}
    \label{fig:SH}
\end{figure}

\subsection{Microphone Configuration}

To capture the sound field with a high spatial resolution, we propose a 64-channel 7th-order HOA framework. Traditional grid-based microphone arrays often suffer from spatial aliasing as mentioned earlier. We mitigate this by placing the microphones using a superposition arrangement, which ensures even sampling of the sound field. The simulation strategy is implemented in the \texttt{Pyroomacoustics} library for the ease of room acoustic simulation and automation. 

However, since \texttt{Pyroomacoustics} does not include Spherical Harmonics directivity, it was implemented from scratch. A new \texttt{SphericalHarmonicDirectivity} class was created in the Pyroomacoustics' directivities framework which takes the order and degree of the microphone as the input and generates the directivity response compatible with the Pyroomacoustics' room simulation framework.

Each microphone's directivity is decomposed into spherical harmonics using the following spherical harmonic function $ Y_n^m(\theta, \phi) $, where $ n $ is the order and $ m $ is the degree:

\begin{equation}
    Y_n^m(\theta, \phi) = \sqrt{\frac{(2n+1)}{4\pi} \cdot \frac{(n-m)!}{(n+m)!}} P_n^m(\cos\theta) e^{im\phi}
\end{equation}

Here, $ P_n^m $ represents the associated Legendre polynomial, $ \theta $ is the zenith angle (or colatitude), and $ \phi $ is the azimuth.

We apply N3D normalization for the spherical harmonic coefficients, which ensures that the spherical harmonics are orthonormal. The normalization factor is:

\begin{equation}
    N_n^m = \sqrt{\frac{(2n+1)}{4\pi} \frac{(n-m)!}{(n+m)!}}
\end{equation}

Using these spherical harmonics, we can decompose the sound field into its HOA components up to the 7th order. For each microphone position $ \mathbf{r}_m $, the signal can be expressed as a weighted sum of the spherical harmonic functions:

\begin{equation}
    p(\mathbf{r}_m, t) = \sum_{n=0}^{N} \sum_{m=-n}^{n} c_n^m(t) Y_n^m(\theta_m, \phi_m)
\end{equation}

where $ c_n^m(t) $ are the HOA coefficients.

\begin{figure}[h]
    \centering
    \includegraphics[width=0.95\columnwidth]{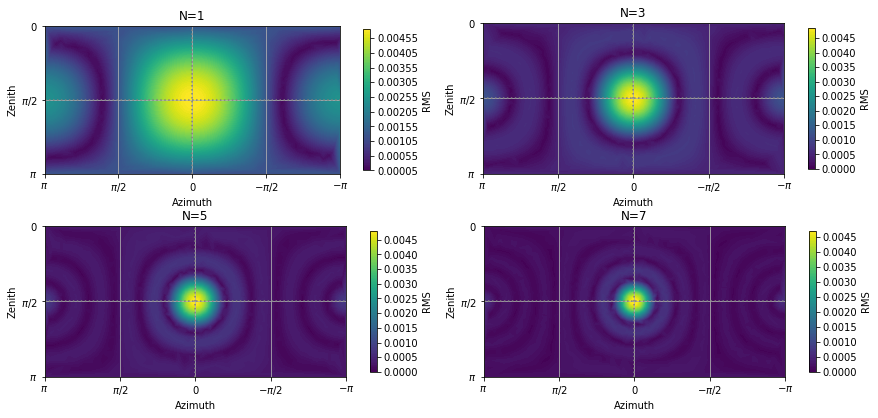}
    \caption{Free field point source a meter away from the microphone array.}
    \label{fig:freefield}
\end{figure}

To allow pyroomacoustics to be able to simulate the RIRs, the Real Spherical Harmonics \( Y_{n,m} \) are derived from the Complex Spherical Harmonics \( Y_n^m \) and are defined as follows:

\begin{equation}
Y_{n,m}(\theta, \phi) = 
\begin{cases} 
\sqrt{2} (-1)^m \text{Im}\left(Y_n^{|m|}(\theta, \phi)\right), & m < 0 \\
Y_n^0(\theta, \phi), & m = 0 \\
\sqrt{2} (-1)^m \text{Re}\left(Y_n^{m}(\theta, \phi)\right), & m > 0
\end{cases}
\end{equation}

where \( \theta \) is the colatitude, \( \phi \) is the azimuth, and \( Y_n^m(\theta, \phi) \) are the complex spherical harmonics $Y_n^m(\theta, \phi)$. The derived microphone directivities are given in Figure \ref{fig:SH} (up to 4th order). The free field measurement of the proposed microphone configuration is given in Figure \ref{fig:freefield}.

\subsection{Room Simulation}

\begin{figure}[!h]
    \centering
    \includegraphics[width=0.9\columnwidth]{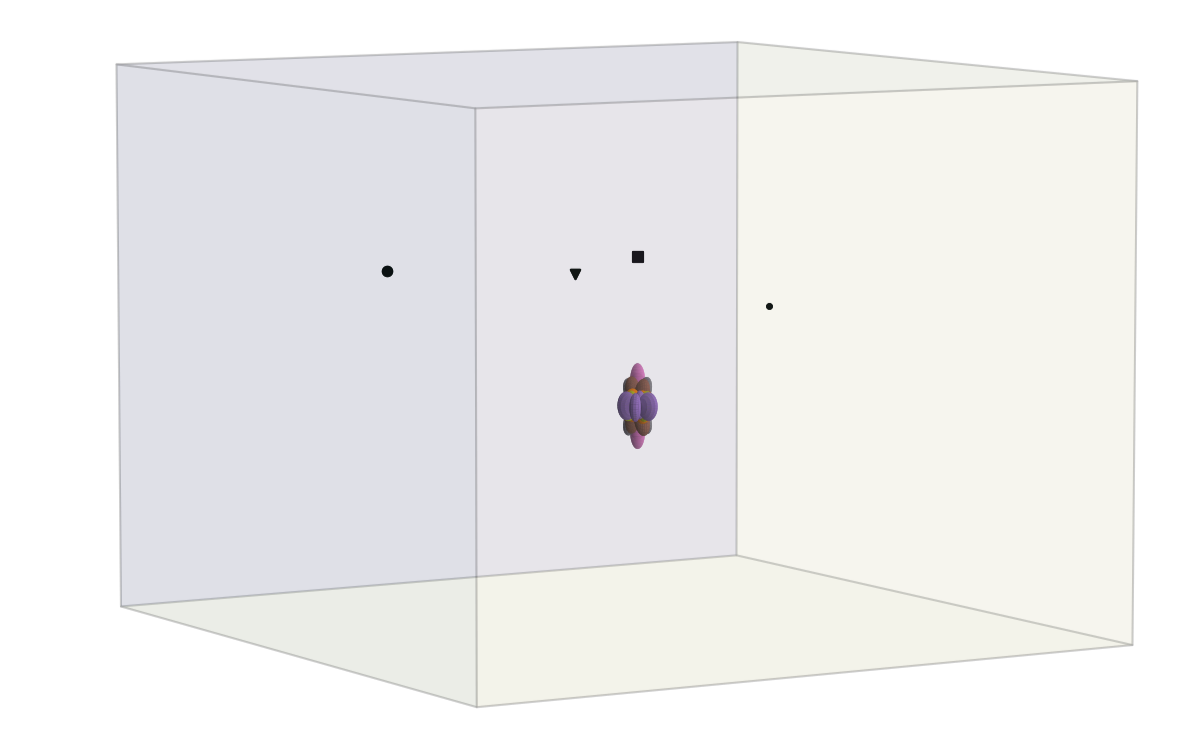}
    \caption{Example room configuration with 4 point sources and the proposed HOA microphone.}
    \label{fig:roomfig}
\end{figure}

Since the \texttt{Pyroomacoustics} library does not support Ray Tracing with microphone directivity, for room acoustic simulations we utilize the Image Source Method (ISM) \cite{allen1979image}. This method models sound reflections in an enclosed space by treating the sound source and its reflections as a set of image sources. Each reflection is computed based on the geometry of the room and the absorption coefficients of its surfaces. We simulate each room with a very image source order (40) and a large variety of room environments, including room geometries, wall absorption coefficients, and source-receiver configurations. A point source and the HOA microphone were placed randomly within the room boundaries, and for each room configuration, the corresponding RIRs were computed for all 64 microphones using the method described. An example simulation can be seen in Figure \ref{fig:roomfig} where 4 different sources and a HOA microphone are placed in a room.

\subsection{Absorption Material Lookup Table}

In simulating realistic room acoustics, selecting appropriate absorption coefficients for materials is essential. A practical approach to assigning acoustic properties is using a predefined lookup table of common materials based on their absorption characteristics across frequency bands. This lookup table spans a wide range of materials typically found on walls, floors, and ceilings, allowing for a more generalized and adaptable configuration compared to direct measurements in specific scenarios, such as curtains on the floor or highly uncommon materials in unrealistic placements.

The lookup table includes materials such as \textit{brickwork}, \textit{ceramic tiles}, and \textit{plasterboard} for walls, while floor materials consist of options like \textit{carpet}, \textit{concrete}, and \textit{marble}. Ceiling materials are selected from absorbers like \textit{fibre panels}, as well as the harder reflective surfaces that are common in various environments. Each material is referenced by its type and associated location, ensuring accurate and realistic RIR generation across various room configurations. This method improves the flexibility of the room models and enhances the reproducibility of results in different acoustic conditions.

This lookup table allows for the realistic simulation of different room configurations, ensuring that the generated RIRs accurately reflect the acoustic properties of common listening environments.

\subsection{Dataset Creation and Ambisonic Encoding}

We simulate various room configurations, including small, medium, and large spaces with different absorption characteristics. The simulation involves:
\begin{itemize}
    \item \textbf{Room Dimensions:} Varied to include cuboid, L-shaped, and hexagonal geometries.
    \item \textbf{Absorption Materials:} Different absorption coefficients are assigned to room surfaces, simulating materials such as carpets, curtains, and reflective surfaces chosen from a lookup table mentioned in the previous section.
    \item \textbf{Source-Receiver Positions:} These are randomized to provide a diverse range of acoustic conditions. Each room consists 20 combinations of source and receiver positions.
\end{itemize}

The final dataset consists of 100,000 simulated 7th-order Ambisonics RIRs, each corresponding to a unique room configuration and source-microphone setup. Each RIR is stored in the AmbiX format (ACN/SN3D),which is widely compatible with spatial audio rendering engines. The metadata associated with each RIR including room dimensions, absorption properties, and source-receiver configurations are provided in the form of a $.csv$ file.  This comprehensive dataset is designed to support a wide range of research in spatial audio, from room acoustics modelling to machine learning-based sound field reconstruction. 

\section{Discussion}

\begin{table*}[t]\centering
\centering
\caption{Comparison of Ambisonic RIR Datasets}
\label{tab:dataset_comparison}
\begin{tabular}{|l|c|c|c|c|}
\hline
\textbf{Dataset}              & \textbf{Ambisonics Order} & \textbf{Number of RIRs} & \textbf{Room Configurations}   \\ \hline
\textbf{HARP (Proposed)}      & 7th                      & 100,000               & Wide variety  \\ \hline
\textbf{C4DM \cite{5496083}}         & 1st                      & 700                    & 3 rooms but multiple positions             \\ \hline
\textbf{TAU-SRIR \cite{politis2020dataset}}   & 4th                     & 114                  & 9 room configurations             \\ \hline
\textbf{MOTUS \cite{gotz2021dataset}}         & 3rd                      & 3,320                    & 830 furniture configuration in a single room        \\ \hline
\textbf{OpenAIR \cite{murphy2010openair}}       & 1st                      & $<$50                    &  Extreme but real scenarios    \\ \hline
\textbf{ARNI SRIR \cite{mckenzie2021datasetspatialroomimpulse}}         & 4th                      & -                    & 5 Room configurations 6-DOF             \\ \hline
\textbf{HOMULA-RIR \cite{HOMULA}}         & 2nd                      & 25                    & Multiple positions in a seminar room        \\ \hline
\end{tabular}
\end{table*}

\begin{figure}[hb]
    \centering
    \includegraphics[width=0.95\columnwidth]{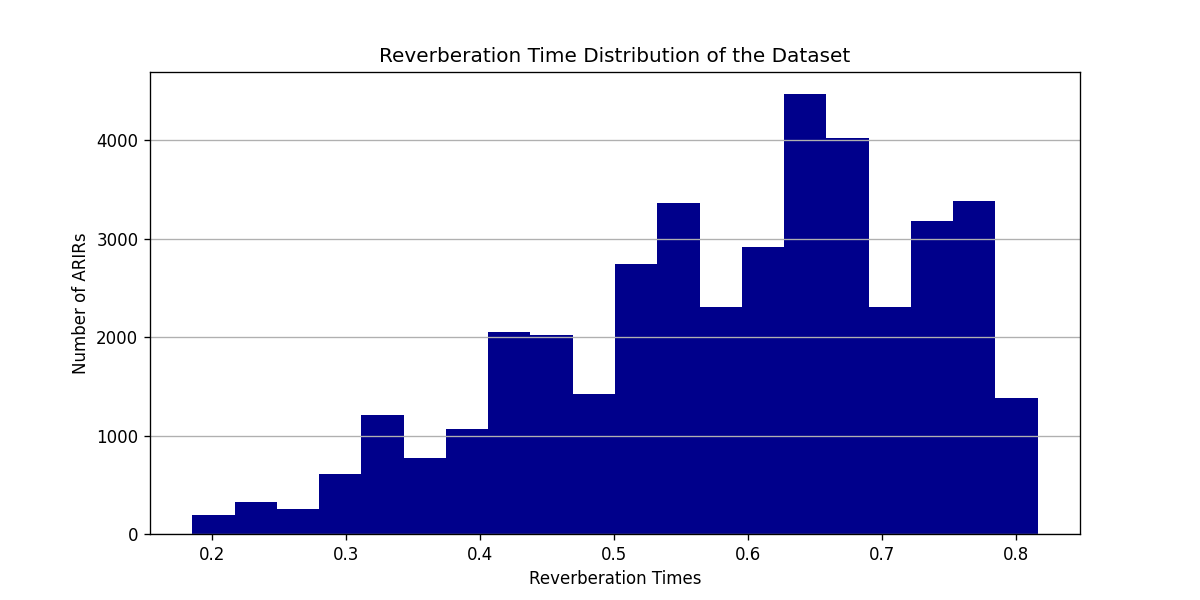}
    \caption{RT60 Distribution calculated with $0^{th}$-order microphone. Histogram to be updated in camera-ready version.}
    \label{fig:RTdist}
\end{figure}

\begin{figure}
    \centering
    \includegraphics[width=0.9\columnwidth]{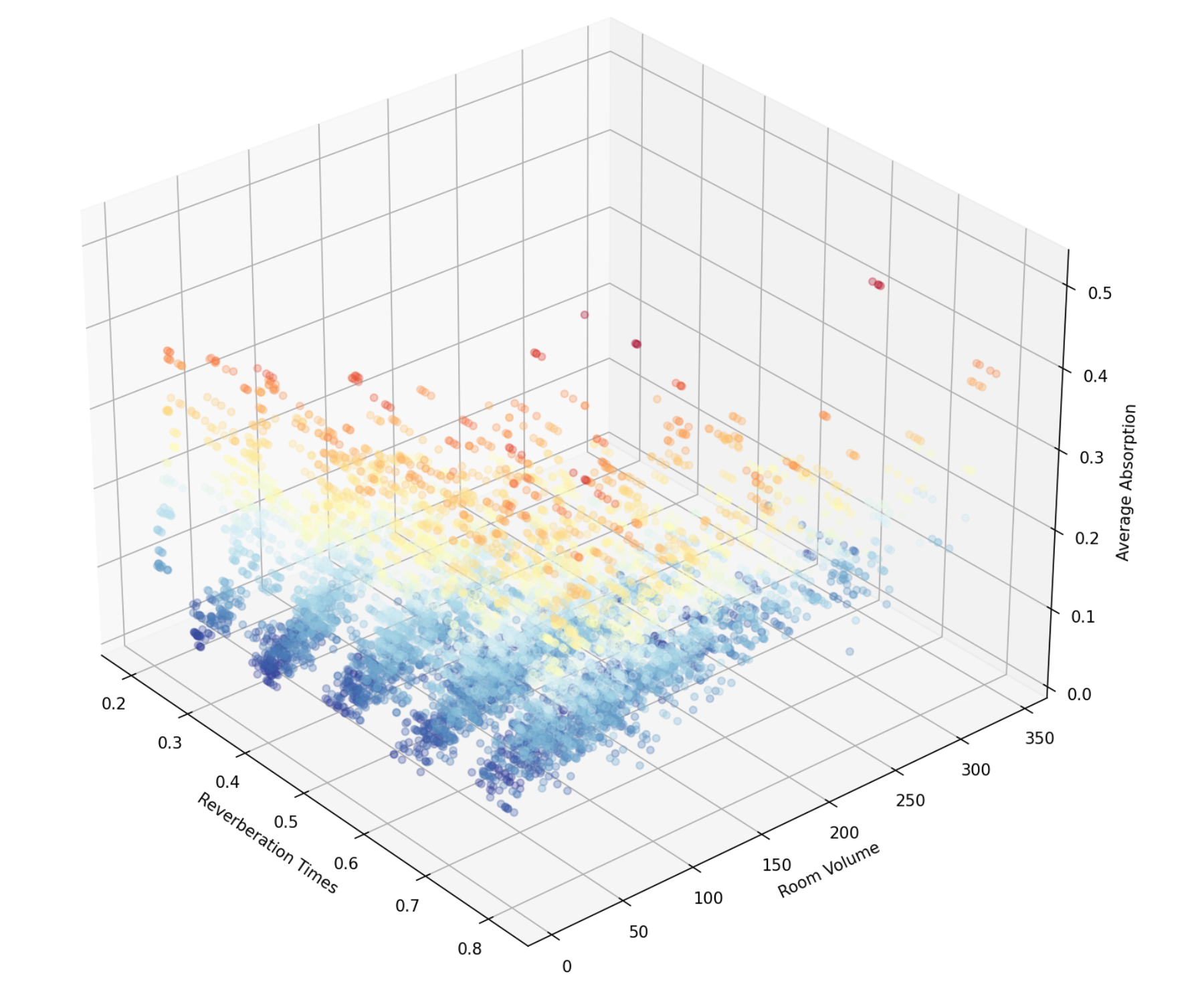}
    \caption{Distribution of the dataset in relation to RT60 [s], Room Volume [$m^3$] and average absorption. Color map (Blue-Red) is applied for better readability of z-axis (Absorption) meaning higher absorption would lead to red color while reflective rooms are marked in blue.}
    \label{fig:distribution}
\end{figure}

A comprehensive analysis of the provided dataset was performed to ensure its usability and practical application in various spatial audio and machine learning tasks. Figure \ref{fig:RTdist} shows the RT60 distribution calculated using the 0th-order (Omnidirectional) microphone. The majority of Room Impulse Responses (RIRs) exhibit RT60 values within the range of 0.4 to 0.8 seconds. This range captures typical room reverberation characteristics in real-world environments, such as living rooms, offices, and lecture halls. These values ensure that the dataset is relevant for a broad range of acoustic conditions encountered in real-world applications. Furthermore, Figure \ref{fig:distribution} summarizes the distribution of RIRs based on RT60 values, room volumes, and average absorption coefficients. As shown, the dataset focuses on typical real-world scenarios, while also including edge cases to support more extreme acoustic conditions. 

Table \ref{tab:dataset_comparison} highlights key differences between HARP and other publicly available Spatial RIR datasets. HARP distinguishes itself by offering 7th-order HOA-RIRs, making it more suitable for advanced spatial audio applications. In contrast, most other datasets are limited to lower orders of ambisonics or provide only a few room configurations, reducing their usefulness for machine learning based approaches.

In contrast, HARP features a significantly larger number of samples—100,000 RIRs, in comparison to other datasets which typically contain only a few hundred RIRs. It also offers greater variety in room configurations and absorption materials, with a practical lookup table tailored for real-world applications. Additionally, the Ambix format ensures compatibility with existing spatial audio frameworks, setting HARP apart as a comprehensive dataset for both academic and industrial research in spatial audio. This level of detail and scale makes the dataset ideal for training machine learning models on spatially aware sound, while most other datasets focus on simpler, less diverse configurations, which limits their generalizability across varied environments.

\section{Conclusion}

This paper presents HARP, a large-scale dataset of up to 7th-order Higher-Order Ambisonic Room Impulse Responses, generated using the Image Source Method and carefully curated to cover a wide variety of acoustic environments. HARP offers an vast amount of spatial RIR data, featuring 100,000 SRIR samples with diverse room configurations, absorption materials, and spatial characteristics. One of the key contributions of this dataset is the creation of a 64 7th-order spherical harmonic microphone configuration based on superposition principle which was used to synthesize the RIRs. 

Despite the high quality and large scale of the HARP dataset, several limitations remain. First, the dataset is generated using the Image Source Method, which, while efficient, may not capture certain real-world acoustic phenomena such as diffraction and scattering with the same precision as other methods like Boundary Element Methods (BEM). Moreover, while the dataset covers a wide range of room types, it could benefit from additional real-world objects such as chairs, tables etc. to simulate a more realistic sound field. 

Future work will focus on expanding the HARP dataset in several directions. First, incorporating simulation of real-world objects to increase the dataset's robustness and provide valuable data for comparing synthetic and real-world acoustics. Additionally, future updates may involve adding dynamic environments where the room configuration or source-receiver positions change over time, catering to emerging research areas in dynamic sound field modelling.

Finally, there is potential for using HARP in the machine learning domain. By using this dataset to train (pre-train) spatial audio models, future work could explore improving source separation, dereverberation, and spatial audio enhancement in real-world interactive systems. In summary, HARP provides an essential foundation for current and future research in spatial audio, room acoustics, and machine learning, with opportunities for further development and refinement.

\bibliography{bibliography} \bibliographystyle{ieeetr}

\begin{thebibliography}{10}

\bibitem{moreau20063d}
S.~Moreau, J.~Daniel, and S.~Bertet, ``3d sound field recording with higher
  order ambisonics--objective measurements and validation of a 4th order
  spherical microphone,'' in {\em 120th Convention of the AES}, pp.~20--23,
  2006.

\bibitem{AudMobNet}
S.~Saini and J.~Peissig, ``Blind room acoustic parameters estimation using
  mobile audio transformer,'' in {\em 2023 IEEE Workshop on Applications of
  Signal Processing to Audio and Acoustics (WASPAA)}, 2023.

\bibitem{bryanCNN}
N.~J. Bryan, ``Impulse response data augmentation and deep neural networks for
  blind room acoustic parameter estimation,'' in {\em 2020 IEEE International
  Conference on Acoustics, Speech and Signal Processing (ICASSP)}, 2020.

\bibitem{CRNN}
S.~Deng, W.~Mack, and E.~A. Habets, ``{Online Blind Reverberation Time
  Estimation Using CRNNs},'' in {\em Proc. Interspeech 2020}, pp.~5061--5065,
  2020.

\bibitem{gotzjournal}
P.~Götz, C.~Tuna, A.~Walther, and E.~A.~P. Habets, ``{Online reverberation
  time and clarity estimation in dynamic acoustic conditions},'' {\em The
  Journal of the Acoustical Society of America}, vol.~153, pp.~3532--3542, 06
  2023.

\bibitem{habets2007single}
E.~A.~P. Habets, ``Single-and multi-microphone speech dereverberation using
  spectral enhancement,'' 2007.

\bibitem{jin2022metamgc}
C.~Jin, F.~Wu, J.~Wang, Y.~Liu, Z.~Guan, and Z.~Han, ``Metamgc: a music
  generation framework for concerts in metaverse,'' {\em EURASIP journal on
  audio, speech, and music processing}, vol.~2022, no.~1, p.~31, 2022.

\bibitem{hold2024}
C.~Hold, ``A parametric spatial audio compression codec for higher-order
  ambisonics,'' 2024.

\bibitem{nilsI3Dparameteric}
N.~Meyer-Kahlen, S.~J. Schlecht, and T.~Lokki, ``Parametric late reverberation
  from broadband directional estimates,'' in {\em 2021 Immersive and 3D Audio:
  from Architecture to Automotive (I3DA)}, pp.~1--10, 2021.

\bibitem{xia2023upmix}
J.~Xia and W.~Zhang, ``Upmix b-format ambisonic room impulse responses using a
  generative model,'' {\em Applied Sciences}, vol.~13, no.~21, p.~11810, 2023.

\bibitem{politis2020overview}
A.~Politis, A.~Mesaros, S.~Adavanne, T.~Heittola, and T.~Virtanen, ``Overview
  and evaluation of sound event localization and detection in dcase 2019,''
  {\em IEEE/ACM Transactions on Audio, Speech, and Language Processing},
  vol.~29, pp.~684--698, 2020.

\bibitem{pyroomacoustics}
R.~Scheibler, E.~Bezzam, and I.~Dokmani{\'c}, ``Pyroomacoustics: A python
  package for audio room simulation and array processing algorithms,'' in {\em
  2018 IEEE international conference on acoustics, speech and signal processing
  (ICASSP)}, pp.~351--355, IEEE, 2018.

\bibitem{zotter2019ambisonics}
F.~Zotter and M.~Frank, {\em Ambisonics: A practical 3D audio theory for
  recording, studio production, sound reinforcement, and virtual reality}.
\newblock Springer Nature, 2019.

\bibitem{frank2014}
M.~Frank, ``How to make ambisonics sound good,'' in {\em Forum
  Acusticum,(Krakow)}, 2014.

\bibitem{bertet2013}
S.~Bertet, J.~Daniel, E.~Parizet, and O.~Warusfel, ``Investigation on
  localisation accuracy for first and higher order ambisonics reproduced sound
  sources,'' {\em Acta Acustica united with Acustica}, vol.~99, no.~4,
  pp.~642--657, 2013.

\bibitem{daniel1998ambisonics}
J.~Daniel, J.-B. Rault, and J.-D. Polack, ``Ambisonics encoding of other audio
  formats for multiple listening conditions,'' in {\em Audio Engineering
  Society Convention 105}, Audio Engineering Society, 1998.

\bibitem{jot1999comparative}
J.-M. Jot, V.~Larcher, and J.-M. Pernaux, ``A comparative study of 3-d audio
  encoding and rendering techniques,'' in {\em Audio engineering society
  conference: 16th international conference: Spatial sound reproduction}, Audio
  Engineering Society, 1999.

\bibitem{allen1979image}
J.~B. Allen and D.~A. Berkley, ``Image method for efficiently simulating
  small-room acoustics,'' {\em The Journal of the Acoustical Society of
  America}, vol.~65, no.~4, pp.~943--950, 1979.

\bibitem{5496083}
R.~Stewart and M.~Sandler, ``Database of omnidirectional and b-format room
  impulse responses,'' in {\em 2010 IEEE International Conference on Acoustics,
  Speech and Signal Processing}, pp.~165--168, 2010.

\bibitem{politis2020dataset}
A.~Politis, S.~Adavanne, and T.~Virtanen, ``A dataset of reverberant spatial
  sound scenes with moving sources for sound event localization and
  detection,'' in {\em Detection and Classification of Acoustic Scenes and
  Events Workshop}, 2020.

\bibitem{gotz2021dataset}
G.~G{\"o}tz, S.~J. Schlecht, and V.~Pulkki, ``A dataset of higher-order
  ambisonic room impulse responses and 3d models measured in a room with
  varying furniture,'' in {\em 2021 Immersive and 3D Audio: from Architecture
  to Automotive (I3DA)}, pp.~1--8, IEEE, 2021.

\bibitem{murphy2010openair}
D.~T. Murphy and S.~Shelley, ``Openair: An interactive auralization web
  resource and database,'' in {\em Audio Engineering Society Convention 129},
  Audio Engineering Society, 2010.

\bibitem{mckenzie2021datasetspatialroomimpulse}
T.~McKenzie, L.~McCormack, and C.~Hold, ``Dataset of spatial room impulse
  responses in a variable acoustics room for six degrees-of-freedom rendering
  and analysis,'' 2021.

\bibitem{HOMULA}
F.~Miotello, P.~Ostan, M.~Pezzoli, L.~Comanducci, A.~Bernardini, F.~Antonacci,
  and A.~Sarti, ``Homula-rir: A room impulse response dataset for
  teleconferencing and spatial audio applications acquired through higher-order
  microphones and uniform linear microphone arrays,'' in {\em 2024 IEEE
  International Conference on Acoustics, Speech, and Signal Processing
  Workshops (ICASSPW)}, pp.~795--799, 2024.

\end{thebibliography}

\vspace{12pt}
\end{document}